\documentclass[a4paper]{article}

\usepackage{INTERSPEECH2019}
\usepackage{diagbox}
\usepackage{caption}
\usepackage{dblfloatfix}
\title{End-to-End Speech Recognition with High-Frame-Rate Features Extraction}
\name{Cong-Thanh Do}
\address{
  Toshiba's Cambridge Research Laboratory, Cambridge, United Kingdom}
\email{cong-thanh.do@crl.toshiba.co.uk}

\begin{document}

\maketitle
\begin{abstract}
  State-of-the-art end-to-end automatic speech recognition (ASR) extracts acoustic features from input speech signal every 10 ms which corresponds to a frame rate of 100 frames/second. In this report, we investigate the use of high-frame-rate features extraction in end-to-end ASR. High frame rates of 200 and 400 frames/second are used in the features extraction and provide additional information for end-to-end ASR. The effectiveness of high-frame-rate features extraction is evaluated independently and in combination with speed perturbation based data augmentation. Experiments performed on two speech corpora, Wall Street Journal (WSJ) and CHiME-5, show that using high-frame-rate features extraction yields improved performance for end-to-end ASR, both independently and in combination with speed perturbation. On WSJ corpus, the relative reduction of word error rate (WER) yielded by high-frame-rate features extraction independently and in combination with speed perturbation are up to 21.3\% and 24.1\%, respectively. On CHiME-5 corpus, the corresponding relative WER reductions are up to 2.8\% and 7.9\%, respectively, on the test data recorded by microphone arrays and up to 11.8\% and 21.2\%, respectively, on the test data recorded by binaural microphones. 
\end{abstract}

\vspace{-5pt}
\section{Introduction}

End-to-end automatic speech recognition (ASR) uses a single neural network architecture within a deep learning framework to perform speech-to-text task \cite{graves2014}. There are two major approaches for end-to-end ASR; attention-based approach uses an attention mechanism to create required alignments between acoustic frames and output symbols which have different lengths, and connectionnist temporal classification (CTC) approach uses Markov assumptions to address sequential problems by dynamic programming \cite{graves2014, watanabe2017}.

In the attention-based end-to-end approach, an encoder-decoder architecture is used to solve the speech-to-text problem which is formulated as a sequence mapping from speech feature sequence to text \cite{chorowski2015, chan2016}. In the encoder-decoder architecture, the input feature vectors are converted into a frame-wise hidden vector by the encoder. In this architecture, bidirectional long short-term memory (BLSTM) \cite{hochreiter1997, graves2014} are often used as an encoder network \cite{watanabe2017}. A pyramid BLSTM (pBLSTM) encoder with subsampling was found to yield better performance than the BLSTM encoder \cite{chan2016}. In \cite{hori2017}, initial layers of the VGG net architecture (deep convolutional neural network (CNN)) \cite{lecun1995, simonyan2015} was found to be helpful when being used prior to the BLSTM in the encoder network. The encoder consisting of the VGG net and the pBLSTM yields better performance than the pBLSTM encoder in many cases \cite{watanabe2018}.

State-of-the-art end-to-end ASR extracts acoustic features from input speech signal every 10 ms which corresponds to a frame rate of 100 frames/second. Extracting acoustic features at frame rates higher than 100 frames/second could gain more information from the input speech signal. The temporal resolution of the feature matrices is also increased and could be useful for end-to-end ASR which uses the VGG net and pBLSTM for encoder because these networks make use of temporal information from input features.

In this report, we investigate the use of high-frame-rate features extraction in end-to-end ASR. High frame rates of 200 and 400 frames/second are used in the features extraction for end-to-end ASR with hybrid CTC/attention architecture \cite{watanabe2017}. The effectiveness of the high-frame-rate features extraction is evaluated independently and in combination with speed perturbation based data augmentation \cite{ko2015}. Experiments are carried out with two speech corpora, the Wall Street Journal (WSJ) corpus \cite{paul1992} and the CHiME-5 corpus which was used for the CHiME 2018 speech separation and recognition challenge \cite{barker2018}. CHiME-5 is a large scale corpus of real multi-speaker conversational speech recorded via multi-microphone hardware in multiple homes. The main difficulty of this corpus comes from the source and microphone distance in addition to the spontaneous and overlapped nature of speech \cite{barker2018}. We show the effectiveness of using high-frame-rate features extraction in end-to-end ASR, independently and in combination with speed perturbation based data augmentation.


\vspace{-10pt}
\section{Related works}
\label{sec:related_works}

\begin{figure*}[!t]
	\centering
		\includegraphics[width=1.8\columnwidth]{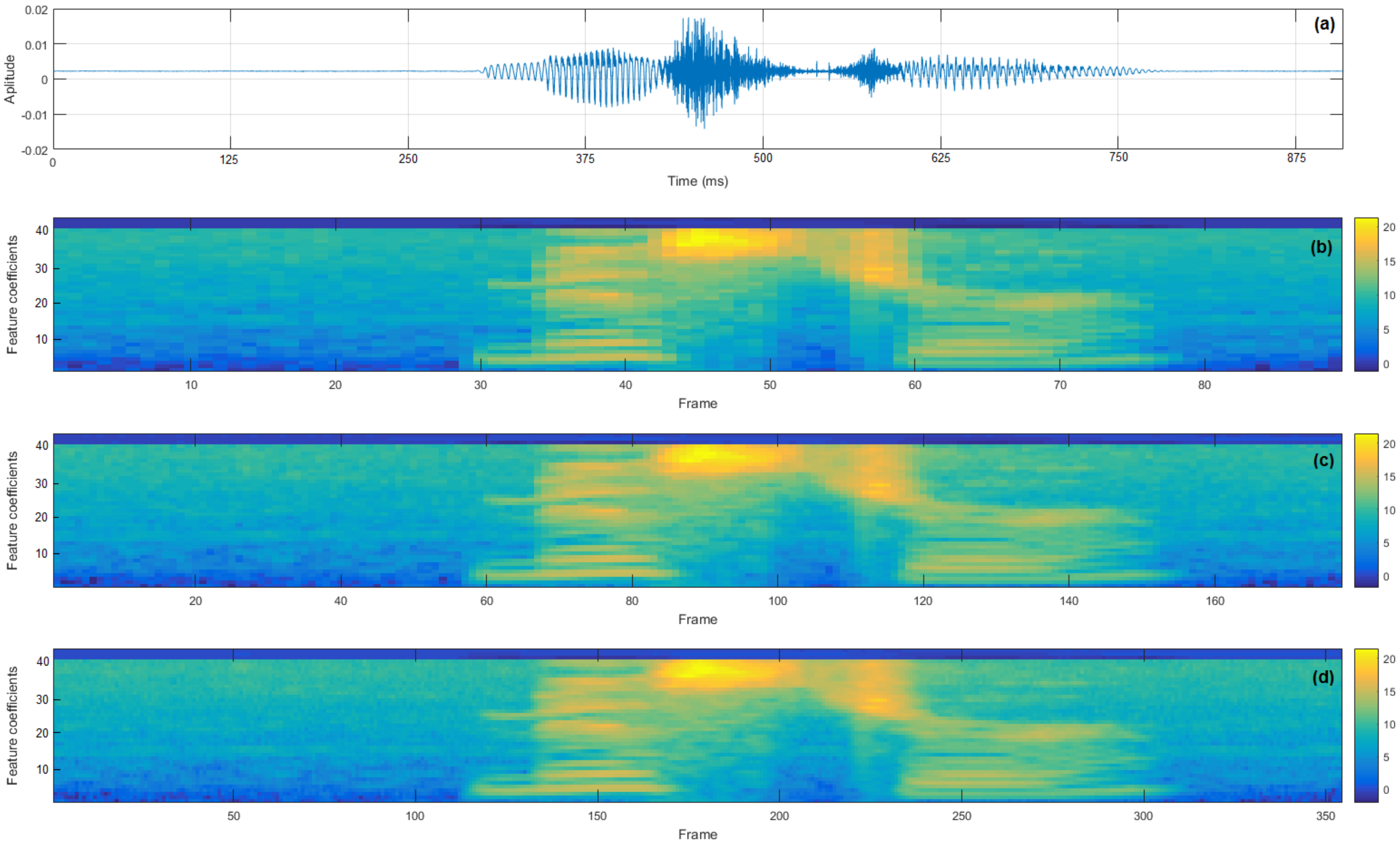}
	\caption{\label{fig:analysis} Unnormalized feature matrices extracted from a speech utterance (a) in the WSJ corpus at frame rates of 100 (b), 200 (c), and 400 (d) frames/second, respectively. Utterance-level mean normalization is applied on the features prior to training and test. \vspace{-10pt}}
\end{figure*}

Variable frame rate analysis was investigated in hidden Markov model (HMM)-Gaussian mixture model (GMM) based ASR \cite{zhu2000, macias2003, tan2010}. In this analysis, frame rates higher than 100 frames/second are used for the rapidly-changing speech segments with relatively high energy while frame rates lower than 100 frames/second are used for steady-state speech segments. The average frame rate is in fact not higher than 100 frames/second. The present report examines features extraction with constant frame rates which are higher than 100 frames/second in end-to-end ASR.

Speed perturbation \cite{ko2015} is a data augmentation technique which creates warped time signals in addition to the original speech signals. Given an audio signal of length $L$ and a warping factor $\alpha$, speed perturbation creates a new signal with duration $\alpha L$ by resampling the original signal with a sampling rate of $\alpha f_s$, where $f_s$ is the sampling rate of the original signal. Speed perturbation shifts the speech spectrum and also results in change in number of frames as the duration of the resulting signal is different \cite{ko2015}. 

Both speed perturbation and high-frame-rate features extraction aim at extracting more information for using in ASR from the original dataset. In fact, high-frame-rate features extraction creates additional speech frames by reducing the hop sizes between adjacent frames whereas speed perturbation creates more data by modifying the length of the signals using resampling. The two approaches of creating more information are thus different and could be complementary to each other.


\section{High-frame-rate features extraction}
\label{sec:analysis}

State-of-the-art end-to-end ASR systems typically extract feature vectors every 10ms which corresponds to a frame rate of 100 frames/second. When high-frame-rate features extraction of 200 and 400 frames/second are used, feature vectors are extracted every 5 and 2.5 ms, respectively. When the hop size is reduced, more feature vectors are extracted and the temporal resolution of the feature matrices is increased. 

In this work, Mel filter-bank (FBANK) features \cite{davis1980, mohamed2012} of 40 dimensions are used. The FBANK features are extracted in a conventional manner as follows: speech signal is first pre-emphasized by using a filter having a transfer function $H[z] = 1 - 0.97z^{-1}$. Speech frames of 25 ms are then extracted at a given frame rate and multiplied with Hamming windows. Discrete Fourier transform (DFT) is used to transform speech frames into spectral domain. Sums of the element-wise multiplication between the magnitude spectrum and the Mel-scale filter-bank are computed. The FBANK coefficients are obtained by taking logarithm of these sums. The FBANK features are augmented with 3-dimensional pitch features which include the value of pitch, delta-pitch and the probability of voicing at each frame \cite{ghahremani2014, barker2018}. In this work, the FBANK and pitch features are extracted using the Kaldi speech recognition toolkit \cite{povey2011}.

Figs. \ref{fig:analysis} (b), \ref{fig:analysis} (c), \ref{fig:analysis} (d) show examples of the 43-dimensional FBANK+pitch feature matrices extracted from a speech utterance (Fig. \ref{fig:analysis} (a)) in the WSJ corpus at frame rates of 100, 200, and 400 frames/second, respectively. It can be observed from these figures that the temporal resolution of the feature matrices increases when the frame rate increases. This higher temporal resolution could provide additional temporal information for the encoder network using VGG net and pBLSTM which make use of the temporal information from input features. ASR experiments are carried out to examine the temporal resolutions of the feature matrices which are useful for end-to-end ASR.

\vspace{-10pt}
\section{Speech corpora}
\label{sec:speech_corpora}

We carry out experiments on two speech corpora, the Wall Street Journal (WSJ) corpus \cite{paul1992} and the CHiME-5 corpus which was used for the CHiME 2018 speech separation and recognition challenge \cite{barker2018}. These two different ASR tasks, one consisting of clean speech recorded by single microphone (WSJ task) and another consisting of conversational speech recorded by both distant microphone arrays and binaural microphones (CHiME-5 task), are suitable for evaluating the effectiveness of high-frame-rate features extraction for end-to-end ASR in different scenarios.

\vspace{-5pt}
\subsection{WSJ corpus}
\label{sec:wsj}

WSJ is a corpus of read speech \cite{paul1992}. The speech utterances in the corpus are quite clean. We use the standard configuration \footnotesize \texttt{train\_si284} \normalsize set for training, \footnotesize \texttt{test\_dev93} \normalsize for validation and \footnotesize \texttt{test\_eval92} \normalsize for test evaluation. The training, development, and evaluation sets consist of 37318, 503, and 333 utterances, respectively. These training, development, and evaluation sets are consistent with the definitions in the Kaldi \cite{povey2011} and ESPnet \cite{watanabe2018} recipes for this corpus.

\vspace{-5pt}
\subsection{CHiME-5 corpus}
\label{sec:chime5}

\subsubsection{Recording scenario}
\label{sec:scenario}

CHiME-5 is the first large-scale corpus of real multi-speaker conversational speech recorded via commercially available multi-microphone hardware in multiple homes \cite{barker2018}. Natural conversational speech from a dinner party of 4 participants was recorded for transcription. Each party was recorded with 6 distant Microsoft Kinect microphone arrays and 4 binaural microphone pairs worn by the participants. There are in total 20 different parties recorded in 20 real homes. This corpus was designed for the CHiME 2018 challenge \cite{barker2018}.

Each party has a minimum duration of 2 hours which composes of three phases, each corresponding to a different location: i) kitchen - preparing the meal in the kitchen area; ii) dining - eating meal in the dining area; iii) living - a post-dinner period in a separate living room area. The participants can move naturally within the home in different locations, but they should stay in each location for at least 30 minutes. There is no constraint on the topic of the conversations. The conversational speech is thus spontaneous.

\vspace{-5pt}
\subsubsection{Audio and transcriptions}
\label{sec:audio_and_transcriptions}

The audio of the parties was recorded with a set of six Microsoft Kinect devices which were strategically placed to capture each conversation by at least two devices in each location. Each Kinect device has a linear array of 4 sample-synchronized microphones and a camera. The audio was also recorded with the Soundman OKM II Classic Studio binaural microphones worn by each participant \cite{barker2018}.  

Manual transcriptions were produced for all the recorded audio. The start and end times and the word sequences of an utterance produced by a speaker are manually obtained by listening to the speaker's binaural recording. These information are used for the same utterance recorded by other recording devices but the start and end times are shifted by an amount that compensates for the asynchonization between devices.

\vspace{-5pt}
\subsubsection{Data for training and test}
\label{sec:training_data}

Training, development and evaluation sets are created from the 20 parties. Data from 16 parties are used for training. The data used for training ASR systems combines both left and right channels of the binaural microphone data and a subset of all Kinect microphone data from 16 parties. In this report, the total amount of speech used in the training set is around 167 hours (the \footnotesize \texttt{data/train\_worn\_u200k} \normalsize set \cite{barker2018}). Each of the development and evaluation sets is created from 2 parties of around 4.5 and 5.2 hours of speech, respectively. The speakers in the training, development and evaluation sets are not overlapped.

For the development and evaluation data, information about the location of the speaker and the reference array are provided. The reference array is chosen to be the one that is situated in the same area. In this work, the results are reported for the single-array track \cite{barker2018} where only the data recorded by the reference array is used for recognition. The results on this corpus in the present report are obtained on the development sets consisting of speech data recorded by the binaural microphones (dev-binaural) and the microphone arrays (dev-array) because the transcriptions of the evaluation set are not publicly available at the time of this submission. Utterances having overlapped speech are not excluded from the training and the development sets. In total, the training set consists of around 318K utterances and each development set consists of around 7.4K utterances. The dev-binaural set consists of only signals from the left channel of the binaural microphones \cite{barker2018, watanabe2018}. 

\vspace{-5pt}
\subsection{Data augmentation}
\label{sec:data_augmentation}

Training data can be augmented to avoid over fitting and improve the robustness of the models \cite{ko2015}. Generally, adding more training data helps improving system's performance. In this work, we apply the speed perturbation based data augmentation technique \cite{ko2015, do2019} to increase the amount of training data of the WSJ and CHiME-5 corpora. The speed perturbation technique creates new training data by resampling the original data. Two additional copies of the original training sets are created by modifying the speed of speech to 90\% and 110\% of the original rate. For each corpus, the whole training set after data augmentation is 3 times larger than the original training set. For CHiME-5, due to the change in the length of the signals after resampling, the start and end times of the speech utterances in the parties are automatically updated by scaling the original start and end times with the resampling rates. For WSJ, this change does not affect the features extraction as the feature vectors are extracted from the whole utterances.

\vspace{-5pt}
\section{Experiments}
\label{sec:experiments}

\subsection{Speech recognition systems}
\label{sec:asr_system}

\subsubsection{Front-end processing}
\label{sec:front-end}

Acoustic features are extracted from the training, development, and evaluation sets for training and testing of ASR systems, on the WSJ and CHiME-5 corpora. Utterance-level mean normalization is applied on the features. 

For WSJ, the FBANK+pitch features are extracted from the whole speech utterances. For CHiME-5, the FBANK+pitch features are extracted from speech utterances which are located in long audio sequences by using the provided start and end times. In the training set, individual speech signals from each microphone in each Kinect microphone array are used directly. In the development set using speech from the reference microphone array, speech signals from four microphones in the microphone array is processed with a weighted delay-and-sum beamformer (BeamformIt \cite{anguera2007}) for enhancement prior to features extraction.

Three frame rates are examined in the features extraction: the conventional frame rate of 100 frames/second and two high frame rates of 200 and 400 frames/second. Speed perturbation is applied only on the training sets whereas high-frame-rate features extraction is applied on the training, development, and evaluation sets.

\vspace{-5pt}
\subsubsection{End-to-end ASR architecture}
\label{sec:system_architecture}

\begin{figure}[ht]
	\centering
		\includegraphics[width=0.9\columnwidth]{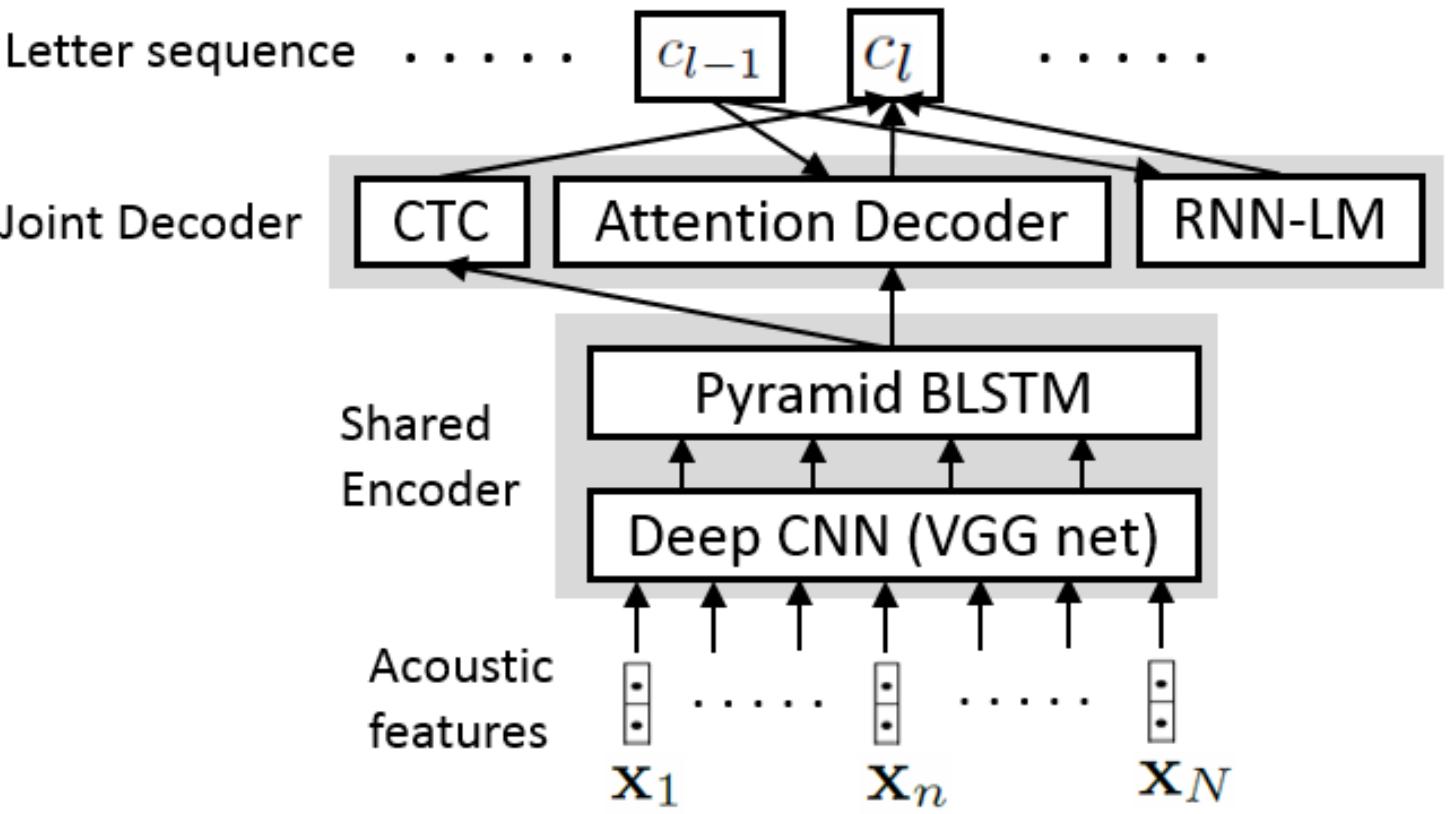}
	\caption{\label{fig:architecture} Hybrid CTC/attention architecture \cite{hori2017, watanabe2017} of the end-to-end ASR systems used in this report. The shared encoder could include either the pBLSTM or the VGG net + pBLSTM.\vspace{-5pt}}
\end{figure}

Hybrid CTC/attention end-to-end ASR systems \cite{watanabe2017} are built using the ESPnet toolkit \cite{watanabe2018}. The system architecture is depicted in Fig. \ref{fig:architecture}. We examine two types of shared encoder, one consists of the initial layers of the VGG net architecture (deep CNN) \cite{lecun1995, simonyan2015} followed by a 4-layer pBLSTM \cite{hochreiter1997, chan2016}, as in \cite{hori2017}, and another consists of the 4-layer pBLSTM. The objective is to examine whether increasing the temporal resolution of the input features could be useful for the VGG net and the pBLSTM which make use of temporal information in the input features.

\vspace{-5pt}

We use a 6-layer CNN architecture which consists of two consecutive 2D convolutional layers followed by one 2D Max-pooling layer, then another two 2D convolutional layers followed by one 2D max-pooling layer. The 2D filters used in the convolutional layers have the same size of 3$\times$3. The max-pooling layers have patch of 3$\times$3 and stride of 2$\times$2. The 4-layer pBLSTM has 320 cells in each layer and direction, and linear projection is followed by each BLSTM layer. The subsampling factor performed by the pBLSTM is 4 \cite{hori2017}. 

In this report, location-based attention mechanism \cite{chorowski2015} is used in the hybrid CTC/attention architecture. This mechanism uses 10 centered convolution filters of width 100 to extract the convolutional features. The decoder network is a 1-layer LSTM with 300 cells. The hybrid CTC/attention architecture is trained within a multi-objective training framework by combining CTC and attention-based cross entropy to improve robustness and achieve fast convergence \cite{watanabe2018}. The training is performed with 15 epochs using the Chainer deep learning toolkit \cite{tokui2015}. The AdaDelta algorithm \cite{zeiler2012} with gradient clipping \cite{pascanu2013} is used for the optimization. We use $\lambda = 0.2$ for WSJ and $\lambda = 0.1$ for CHiME-5 in the multi-objective learning framework for training the hybrid CTC/attention systems \cite{watanabe2017}, in consistent with the ESPnet training recipes for these corpora \cite{watanabe2018}.

During joint decoding, CTC and attention-based scores are combined in a one-pass beam search algorithm \cite{watanabe2018}. A recurrent neural network language model (RNN-LM), which is a 1-layer LSTM, is trained on the transcriptions of the training data, for each corpus. This RNN-LM is used in the joint decoding where its log probability is combined with the CTC and attention scores \cite{watanabe2018}. The weight of the RNN-LM's log probability is set to 0.1 and the beam width is set to 20 during decoding.

\vspace{-5pt}
\subsection{Experimental results}
\label{sec:experimental_results}

Tabs. \ref{tab:WER_WSJ_blstmp} and \ref{tab:WER_WSJ_vggblstmp} show the results in terms of word error rates (WERs) on the WSJ corpus, for the systems using the pBLSTM and the VGG net + pBLSTM encoders, respectively. Tabs. \ref{tab:WER_chime5_blstmp} and \ref{tab:WER_chime5_vggblstmp} show the corresponding WERs on the CHiME-5 corpus. In these tables, the results with ``+SP'' are obtained when speed perturbation (SP) is used to augment the training sets. On both corpora, using features extraction with frame rates higher than the conventional frame rate of 100 frames/second appears to be helpful in reducing the WERs, both when the pBLSTM and the VGG net + pBLSTM encoders are used. Also, high-frame-rate features extraction and speed perturbation based data augmentation are complementary because the gains obtained when using these two methods together are higher than those obtained with each method when they are used separately. In addition, the systems using the VGG net + pBLSTM encoder have lower WERs than those using the pBLSTM encoder, on both corpora.

\vspace{-5pt}
\subsubsection{WSJ}


On WSJ, increasing the frame rate from 200 to 400 frames/second still yields a little WER reduction, but not always. When the VGG net + pBLSTM encoder is used (see Tab. \ref{tab:WER_WSJ_vggblstmp}), the best relative WER reductions on the Dev93 and Eval92 sets obtained with high-frame-rate features extraction are 21.3\% and 12.1\%, respectively. When using high-frame-rate features extraction with speed perturbation, the best relative WER reductions on the Dev93 and Eval92 sets are 24.1\% and 15.1\%, respectively.   

\begin{small}
\begin{table}[t]
\centering
\captionof{table}{WERs on WSJ when the pBLSTM encoder is used.}
\label{tab:WER_WSJ_blstmp}
\begin{tabular}{ |p{2.3cm}|p{0.5cm}|p{0.5cm}|p{0.5cm}|}
 \hline
\diagbox[width=2.72cm]{\footnotesize Test set}{\footnotesize Frame rate} & \footnotesize 100 & \footnotesize 200 & \footnotesize 400 
 \\\hline\hline
 \footnotesize Dev93 & \footnotesize 11.0 & \footnotesize 9.7 & \footnotesize 9.2\\
\hline
 \footnotesize Eval92 & \footnotesize 8.0 & \footnotesize 6.5 & \footnotesize 6.5\\
\hline\hline
 \footnotesize Dev93 (+SP) & \footnotesize 9.9 & \footnotesize 8.9 & \footnotesize 8.9\\
\hline
 \footnotesize Eval92 (+SP) & \footnotesize 7.2  & \footnotesize 6.0 & \footnotesize 6.1\\
\hline
\end{tabular}
\centering
\captionof{table}{WERs on WSJ when the VGG net + pBLSTM encoder is used.}
\label{tab:WER_WSJ_vggblstmp}
\begin{tabular}{ |p{2.3cm}|p{0.5cm}|p{0.5cm}|p{0.5cm}|}
 \hline
\diagbox[width=2.72cm]{\footnotesize Test set}{\footnotesize Frame rate} & \footnotesize 100 & \footnotesize 200 & \footnotesize 400 
 \\\hline\hline
 \footnotesize Dev93 & \footnotesize 10.8 & \footnotesize 8.9 & \footnotesize 8.5\\
\hline
 \footnotesize Eval92 & \footnotesize 6.6 & \footnotesize 6.2 & \footnotesize 5.8\\
\hline\hline
 \footnotesize Dev93 (+SP) & \footnotesize 9.9 & \footnotesize 8.3 & \footnotesize 8.2\\
\hline
 \footnotesize Eval92 (+SP) & \footnotesize 6.8  & \footnotesize 5.6 & \footnotesize 5.8\\
\hline
\end{tabular}
\vspace{2pt}
\centering
\captionof{table}{WERs on CHiME-5 when the pBLSTM encoder is used.}
\label{tab:WER_chime5_blstmp}
\begin{tabular}{ |p{2.3cm}|p{0.5cm}|p{0.5cm}|p{0.5cm}|}
 \hline
\diagbox[width=2.72cm]{\footnotesize Test set}{\footnotesize Frame rate} & \footnotesize 100 & \footnotesize 200 & \footnotesize 400 
 \\\hline\hline
 \footnotesize Dev-binaural & \footnotesize 64.7 & \footnotesize 59.5 & \footnotesize 59.5\\
\hline
 \footnotesize Dev-array & \footnotesize 93.9 & \footnotesize 90.6 & \footnotesize 92.3\\
\hline\hline
 \footnotesize Dev-binaural (+SP) & \footnotesize 58.5 & \footnotesize 50.8 & \footnotesize 51.5\\
\hline
 \footnotesize Dev-array (+SP) & \footnotesize 90.1  & \footnotesize 86.1 & \footnotesize 85.9\\
\hline
\end{tabular}
\centering
\captionof{table}{WERs on CHiME-5 when the VGG net + pBLSTM encoder is used.}
\label{tab:WER_chime5_vggblstmp}
\begin{tabular}{ |p{2.3cm}|p{0.5cm}|p{0.5cm}|p{0.5cm}|}
 \hline
\diagbox[width=2.72cm]{\footnotesize Test set}{\footnotesize Frame rate} & \footnotesize 100 & \footnotesize 200 & \footnotesize 400 
 \\\hline\hline
 \footnotesize Dev-binaural & \footnotesize 61.1 & \footnotesize 53.9 & \footnotesize 53.5\\
\hline
 \footnotesize Dev-array & \footnotesize 89.6 & \footnotesize 87.1 & \footnotesize 87.6\\
\hline\hline
 \footnotesize Dev-binaural (+SP) & \footnotesize 54.6 & \footnotesize 48.1 & \footnotesize 53.3\\
\hline
 \footnotesize Dev-array (+SP) & \footnotesize 85.1 & \footnotesize 82.5 & \footnotesize 88.0\\
\hline
\end{tabular}
\end{table}
\end{small}

\vspace{-5pt}
\subsubsection{CHiME-5}

CHiME-5 is a challenging task with high WERs on the development sets. On this corpus, increasing the frame rate from 200 to 400 frames/second generally does not yield WER reduction. When the VGG net + pBLSTM encoder is used (see Tab. \ref{tab:WER_chime5_vggblstmp}), the baseline system has WERs of 61.1\% and 89.6\% on the dev-binaural and dev-array sets, respectively. The WERs of the baseline system introduced by the challenge organizers on the same sets were 67.2\% and 94.7\%, respectively \cite{barker2018}. In the architecture using the VGG net + pBLSTM encoder, the best relative WER reductions on the dev-binaural and dev-array sets obtained with high-frame-rate features extraction are 11.8\% and 2.8\%, respectively. When using high-frame-rate features extraction in combination with speed perturbation, the relative WER reductions on the dev-binaural and dev-array sets are 21.2\% and 7.9\%, respectively.

\vspace{-5pt}
\section{Conclusion}
\label{sec:conclusion}

This report investigated the use of high-frame-rate features extraction in end-to-end speech recognition. Experimental results on the WSJ and CHiME-5 corpora showed that improved ASR performance was achieved when using features extraction at a frame rate higher than 100 frames/second. These results showed that end-to-end ASR using pBLSTM and VGG net + pBLSTM encoders can make use of additional information from input feature matrices of higher temporal resolution than those extracted with the conventional 100 frames/second frame rate. Using high-frame-rate features extraction in combination with speed perturbation based data augmentation yielded complementary gains. The relative WER reductions obtained by the combination of these two methods were up to 24.1\% and 21.2\% on the WSJ and CHiME-5 corpora, respectively.

\vfill\pagebreak

\bibliographystyle{IEEEtran}

\bibliography{mybib}


\end{document}